\documentclass[prl,twocolumn,nofootinbib,aps,superscriptaddress,tightenlines,preprintnumbers]{revtex4}

\usepackage{epsfig,latexsym,cancel,amssymb,amsmath,verbatim,mathrsfs}
\usepackage{color}
\usepackage{graphicx}

\newcommand{\be}{\begin{equation}}
\newcommand{\ee}{\end{equation}}
\newcommand{\bea}{\begin{eqnarray}}
\newcommand{\eea}{\end{eqnarray}}
\newcommand{\ba}{\begin{array}}
\newcommand{\ea}{\end{array}}

\long\def\symbolfootnote[#1]#2{\begingroup%
\def\thefootnote{\fnsymbol{footnote}}\footnote[#1]{#2}\endgroup} 

\newcommand{\eq}[1]{Eq.~\eqref{#1}}

\newcommand{\beq}{\begin{equation}}
\newcommand{\eeq}{\end{equation}}
\newcommand{\lrf}[2]{\left(\frac{#1}{#2}\right)}

\newcommand{\tev}{\, {\rm TeV}}
\newcommand{\gev}{\, {\rm GeV}}
\newcommand{\mev}{\, {\rm MeV}}
\newcommand{\nnmb}{\nonumber}

\begin{document}


\title{Hylogenesis: \\A Unified Origin for Baryonic Visible Matter and Antibaryonic Dark Matter
} %

\author{Hooman Davoudiasl}
\affiliation{Department of Physics, Brookhaven National Laboratory, Upton, 
NY 11973, USA}
\author{David E. Morrissey} 
\affiliation{Theory Group, TRIUMF, 4004 Wesbrook Mall, Vancouver, BC, 
V6T 2A3, Canada}
\author{Kris Sigurdson} 
\affiliation{Department of Physics and Astronomy, University of British 
Columbia, Vancouver, BC V6T 1Z1, Canada}
\author{Sean Tulin} 
\affiliation{Theory Group, TRIUMF, 4004 Wesbrook Mall, Vancouver, BC, 
V6T 2A3, Canada}

\date{\today}

\begin{abstract}

We present a novel mechanism for generating both the baryon 
and dark matter densities of the Universe.  A new Dirac fermion
$X$ carrying a conserved baryon number charge couples to  
the Standard Model quarks as well as a GeV-scale hidden sector.  
CP-violating decays of $X$, produced non-thermally in low-temperature 
reheating, sequester antibaryon number in the hidden sector, thereby leaving a 
baryon excess in the visible sector.  The antibaryonic hidden states 
are stable dark matter.  
A spectacular signature of this mechanism is the baryon-destroying 
inelastic scattering of dark matter that can annihilate baryons at 
appreciable rates relevant for nucleon decay searches.

\end{abstract}

\maketitle

\noindent {\bf{I. Introduction:}}
  Precision cosmological measurements indicate that a fraction 
$\Omega_b \simeq 0.046$ of the energy content of the Universe consists 
of baryonic matter, while $\Omega_d \simeq 0.23$ is made
up of dark matter (DM)~\cite{Komatsu:2008hk}.  Unfortunately, 
our present understanding 
of elementary particles and interactions, the Standard Model~(SM), 
cannot account for the abundance of either observed component of 
non-relativistic particles.  

  In this \emph{Letter} we propose a unified mechanism, 
\mbox{hylogenesis}\footnote{From Greek, \emph{hyle} ``primordial matter" + \emph{genesis} ``origin."},
to generate 
the baryon asymmetry and the dark matter density simultaneously.  
The SM is extended to include a new hidden sector of states
with masses near a GeV and very weak couplings to the SM. 
Such sectors arise in many well-motivated theories of physics
beyond the SM, and have received much attention within the contexts  
dark matter models~\cite{Pospelov:2008zw}, and high luminosity, 
low-energy precision measurements~\cite{Bjorken:2009mm}.

  The main idea underlying our mechanism is that some of the particles 
in the hidden sector are charged under a generalization
of the global baryon number~($B$) symmetry of the SM.  This symmetry
is not violated by any of the relevant interactions in our model.
Instead, equal and opposite baryon asymmetries are created in the
visible and hidden sectors, and the Universe has zero total $B$.  
These asymmetries are generated when (i)
the TeV-scale states $X_1$ and its antiparticle $\bar{X}_1$ 
(carrying equal and opposite $B$ charge) are generated non-thermally 
in the early Universe (e.g., during reheating), and (ii) $X_1$ decays into
visible and hidden baryonic states.  The $X_1$ decays violate quark 
baryon number and CP, and occur away from equilibrium. 
Both the visible and hidden baryons are stable due to a combination 
of kinematics and symmetries.  The relic density of the hidden baryons 
is set by their asymmetry, and they make up the dark matter of the 
Universe.  We compute the baryon and dark matter densities within 
a concrete model realizing this mechanism in Section II.

  A potentially spectacular signature of our model is that rare 
processes can transfer baryon number from the hidden to the visible sector.  
Effectively, antibaryonic dark matter states can annihilate baryons 
in the visible sector through inelastic scattering.   These events mimic 
nucleon decay into a meson and a neutrino, 
but are distinguishable from standard nucleon decay by the kinematics of 
the meson.  In Section III, we discuss this signature in more detail, 
along with its implications for direct detection and astrophysical systems.

  We note that our scenario shares some
elements with Refs.~\cite{Kitano:2004sv,Kitano:2005ge,An:2009vq,
Allahverdi:2010im,Shelton:2010ta,Haba:2010bm,Agashe:2004bm,Farrar:2005zd}, but involves a different production 
mechanism and unique phenomenological consequences. 


\noindent{\bf{II. Genesis of Baryons and DM\label{sec:asym}:}}
  In our model, the hidden sector consists of
two massive Dirac fermions $X_a$ ($a=1,2$, 
with masses $m_{X_2} > m_{X_1} \gtrsim$ TeV), a Dirac fermion $Y$, 
and a complex scalar $\Phi$ (with masses $m_Y \sim m_\Phi \sim$ GeV).
These fields couple through the ``neutron portal'' ($XU^cD^cD^c$) and
a Yukawa interaction:
\beq
-\mathscr{L} \supset 
\frac{\lambda_a}{M^2}\,\bar{X}_aP_Rd\;\bar{u}^cP_Rd
+ \zeta_a \, \bar{X}_a Y^c \Phi^* + \textrm{h.c.} 
\label{eq:couplings}
\eeq 
Many variations on these operators exist, corresponding to different 
combinations of quark flavors and spinor contractions.  With this set of 
interactions one can define a generalized global baryon number 
symmetry that is conserved, with charges $B_X = -(B_Y + B_\Phi) = 1$.  
The proton, $Y$, and $\Phi$ are stable due to their 
$B$ and gauge charges if their masses satisfy
\beq
|m_Y - m_{\Phi}| < m_p + m_e < m_Y + m_{\Phi} \; .
\label{eq:massdiff}
\eeq
$Y$ and $\Phi$ are the ``hidden antibaryons'' that comprise 
the dark matter.  
Furthermore, there exists a physical CP-violating phase 
$\arg(\lambda_1^* \lambda_2 \zeta_1 \zeta_2^*)$ that cannot be 
removed through phase redefinitions of the fields.

  We also introduce a hidden $U(1)^\prime$ gauge symmetry under 
which $Y$ and $\Phi$ have opposite charges $\pm e^\prime$, while $X_a$ is neutral.  
We assume this symmetry is spontaneously broken at the GeV scale, 
and has a kinetic mixing with SM hypercharge $U(1)_Y$ via the coupling
$-\frac{\kappa}{2}B_{\mu\nu}Z'_{\mu\nu}$,
where $B_{\mu \nu}$ and $Z^\prime_{\mu\nu}$ are the $U(1)_Y$ and $U(1)'$ 
field strength tensors.  At energies well below the electroweak scale 
the effect of this mixing is primarily to generate a vector coupling of the 
massive $Z'$ gauge boson to SM particles with strength 
$-c_W\kappa\,Q_{em}e$.  The $\gev$-scale $Z'$ masses
we consider here can be consistent with observations for 
$10^{-6} \lesssim \kappa \lesssim 10^{-2}$~~\cite{Bjorken:2009mm}.

\begin{figure}[ttt]
\begin{center}
\includegraphics[scale=0.95]{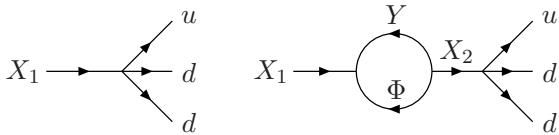}
\end{center}
\vspace{-.4cm}
\caption{Tree-level and one-loop graphs for decay $X_1 \to udd$.  
}
\label{feynman}
\vspace{-0.5cm}
\end{figure}

Baryogenesis begins when a non-thermal, 
CP-symmetric population of $X_1$ and $\bar{X_1}$ is produced in the 
early Universe.  These states decay through 
$X_1 \to udd$
or 
$X_1 \to \bar{Y}\Phi^*$ 
(and their conjugates).  
An asymmetry between the partial widths for 
$X_1 \to udd $ and $\bar{X}_1\to \bar u \bar d \bar d$ 
arises from interference
between the diagrams shown in Fig.~\ref{feynman}, 
and is characterized by
\bea
\epsilon &=& \frac{1}{2\Gamma_{X_1}}\left[\Gamma({X}_1 \to udd)
-\Gamma(\bar{X}_1\to \bar{u}\bar{d}\bar{d})\right]\\
&\simeq& 
\frac{ m_{X_1}^5 
\textrm{Im}[\lambda_1^* \lambda_2 \zeta_1 \zeta_2^*] }
{256 \pi^3 \, |\zeta_1|^2 \, M^4 m_{X_2}} \;,
\nnmb
\eea
where we have assumed that the total decay rate $\Gamma_{X_1}$ is dominated 
by $X_1\to \bar{Y}\Phi^*$ over the three-quark mode, and that $m_{X_2} \gg m_{X_1}$.  
For $\epsilon \ne 0$, $X_1$ decays generate a baryon asymmetry 
in the visible sector, and by CPT an equal and opposite baryon 
asymmetry in the hidden sector.  These asymmetries can be ``frozen in'' 
by the weakness of the coupling between both sectors.  

   We model the non-thermal production of $X_1$ as a reheating process 
after a period where the energy content of the Universe was dominated by 
the coherent oscillations of a scalar field $\varphi$.    
This field could be the inflaton, or it could be a {moduli} field
arising from an underlying theory with supersymmetry~\cite{Coughlan:1983ci}
or a compactification of string theory~\cite{Acharya:2010af}.  
As $\varphi$ oscillates, it decays to visible and hidden sector states 
reheating these two sectors.  
We suppose that a fraction of the $\varphi$ 
energy density $\rho_\varphi$ is converted into $X_1, \bar{X}_1$ states, 
while the remainder goes into visible and hidden sector radiation 
which quickly thermalizes due to gauge interactions.

The dynamics of hylogenesis and reheating are governed by 
the Boltzmann equations
\begin{subequations} \label{eq:boltz}
\bea
&& \,\frac{d}{dt}\!\left(a^3 \rho_\varphi \right)  = 
- \, \Gamma_\varphi \, a^3\rho_\varphi \label{eq:rhophi} \, , \\
%
&& \,\,\,\frac{d}{dt}\!\left(a^3 s \right) =
+ \, \Gamma_\varphi \, a^3 \rho_\varphi/T \label{eq:rhor} \, , \\
%
%
&& \frac{d}{dt}\!\left(a^3 {n}_B\right)  =
\epsilon \,  \mathcal{N}_X \Gamma_\varphi a^3\rho_\varphi/m_\varphi  \; 
\label{eq:nb}
\eea
\end{subequations}
%
%
%
%
%
with $\varphi$ mass $m_\varphi$ and decay rate $\Gamma_\varphi$.
$s \equiv s_{\textrm{HS}} + s_{\textrm{SM}} = (2\pi^2/45)g_s T^3$ is the total entropy density of SM and HS states (assumed in kinetic equilibrium at temperature $T$ with an effective number of entropy degrees of freedom $g_s(T)$\,), and $n_B$ is the baryon number density in the 
visible sector (i.e. quarks).   
The scale factor $a(t)$ is determined by the Friedmann equation 
$H^2 \equiv (\dot{a}/a)^2 = (8\pi G/3) \, (\rho_\varphi+\rho_r)$, where $ \rho_r \equiv (\pi^2/30) g T^4$ is the total radiation density and $g(T)$ is the effective number of degrees of freedom.  $\mathcal{N}_X$ is the average number 
of $X_1$ states produced per $\varphi$ decay.

\eq{eq:rhophi} describes the depletion of the oscillating field 
energy due to redshifting and direct $\varphi$ decays and has the simple solution $\rho_{\varphi} \propto e^{-\Gamma_{\varphi} t} a^{-3}$,
while \eq{eq:rhor}
gives the rate of entropy production due to decays and describes the reheating of the Universe.  We adopt the convention that reheating occurs at temperature $T_{RH}$, defined when $\rho_r(T_{RH}) = \rho_\varphi(T_{RH})$.  This occurs near the characteristic decay time $t \simeq \Gamma_{\varphi}^{-1}$, where the total decay width $\Gamma_{\varphi}$ takes the form~\cite{Allahverdi:2010im,Acharya:2010af} $\Gamma_{\varphi} = {m_{\varphi}^3}/{(4\pi \, \Lambda^2)}$.
Here, $\Lambda$ is a large energy scale corresponding to 
the underlying ultraviolet dynamics.  For example, 
$\Lambda \sim M_\text{Pl} = 2.43\times 10^{18}\,\gev$
for many moduli in string theory or supergravity.  
At reheating, the radiation temperature is approximately~\cite{Allahverdi:2010im}
\bea
T_{RH} 
%
&\simeq& 5\,\mev\,\lrf{10}{g}^{1/4}\lrf{M_\text{Pl}}{\Lambda}
\lrf{m_{\varphi}}{100\,\tev}^{3/2}. \;
\eea
We require $T_{RH} \gtrsim 5\mev$ to
maintain successful nucleosynthesis.

 \eq{eq:nb} determines the comoving density of visible baryons. 
The remnant of the intermediate $X_1$ stage appears in the right-hand-side
of \eq{eq:nb}.  The factor $\epsilon$ encodes the $X_1$ decay asymmetry.
In writing Eq.~\eqref{eq:boltz} we implicitly take 
$m_{X_1} \gg T$ and $\Gamma_{X_1} \gg \Gamma_{\varphi}, H$. 
The former condition implies inverse decays and scattering reactions that could wash out the asymmetry,
such as $\bar{u}X_1\to {d}{d}$ ,
are suppressed by Boltzmann factors of $e^{-m_{X_1}/T}$, while the latter condition is satisified for 
$|\zeta_1| \gg m_{\varphi}^2/(m_{X_1} \Lambda)$. 
The hidden-visible baryon asymmetry can also be washed out by
$Y\Phi \to 3\bar{q}$ scattering.  A sufficient condition for this
washout process to be ineffective is
\beq
T_{RH} \lesssim (2\,\gev)\,
\left(\sum_{a,b}\frac{\lambda_{a}\lambda_{b}^*\zeta_a^*\zeta_b~\tev^6}
{M^4 m_{X_a} m_{X_b}}\right)^{-1/5}.
\eeq
The allowed $T_{RH}$ increases roughly linearly with the mass scale $(M^4 m_{X_{1,2}}^2)^{1/6}$.

The resulting baryon asymmetry today is given by
\beq
\eta_B \equiv {n_B}/{s} = \frac{\epsilon\,\mathcal{N}_X T_{RH}}{m_\varphi} \, f(m_{\varphi} \Gamma_\varphi) \; .
\eeq
Assuming that reheating occurs instantaneously, one can show analytically that $f= 3/4$.  A numerical solution to Eqs.~\eqref{eq:boltz} reveals $f \simeq 1.2$, with less than 10\% variation over a wide range of $(m_\varphi, \Gamma_\varphi)$.
Larger values of $T_{RH}$ (larger $m_{\varphi}$ for fixed $\Lambda$) allow for greater production of baryons.

For the parameter values $m_\varphi=2000$ TeV, $\Lambda=M_{\text{Pl}}$, $\mathcal{N}_X = 1$, we find $T_{RH} \simeq 400$ MeV and $\eta_B/\epsilon \simeq 2.5 \times 10^{-7}$.  The observed value of the baryon asymmetry is obtained for 
$ \textrm{Im}[\lambda_1^* \lambda_2 \zeta_2/\zeta_1] m_{X_1}^5/(M^4 m_{X_2}) \sim 3 $.  Smaller values of $\epsilon$ and $m_\varphi$ are viable for $\Lambda < M_{\textrm{Pl}}$.

  We have implicitly assumed that the $Z^\prime$ maintains 
kinetic equilibrium between the SM and hidden sectors. 
This will occur if $\Gamma_{Z'}\langle 1/\gamma\rangle > H$,
where $\gamma$ is a relativistic time dilation factor, which implies
~\cite{Pospelov:2007mp}
\beq
\kappa > 1.5\times 10^{-8}\,\lrf{g}{10}^{1/2} \lrf{m_{Z'}}{\gev}^{-1} \lrf{T}{\gev}^{3/2} \, ,
\eeq
provided $T_{RH} > m_{Z^\prime}$.
After baryogenesis, the CP-symmetric densities of hidden states are 
depleted very efficiently through annihilation $Y\bar{Y} \to Z^\prime Z^\prime$
and  $\Phi\Phi^* \to Z^\prime Z^\prime$ provided $m_{Z'} < m_Y,\,m_{\Phi}$, 
with the $Z^\prime$ decaying later to SM states by mixing with the photon. 
The cross-section for $Y\bar{Y}\to Z'Z'$ is
given by~\cite{Pospelov:2007mp}
\bea
\langle \sigma v\rangle &=& \frac{{e^\prime}^4}{16\pi}\frac{1}{m_Y^2}
\sqrt{1-m_{Z'}^2/m_Y^2}\\
&\simeq& (1.6\times 10^{-25}cm^3/s)\lrf{e^\prime}{0.05}^4\lrf{3\,\gev}{m_Y}^2.
\nnmb
\eea
Annihilation of $\Phi^*\Phi$ is given by a similar expression.
These cross sections are much larger than what is needed to obtain
the correct DM abundance by ordinary thermal freeze-out, and all of the 
non-asymmetric DM density will be eliminated up to an exponentially small 
remainder~\cite{Griest:1986yu}.
Note that the annihilation process may occur later than is typical
for thermal freeze-out for $T_{RH} \lesssim m_{Y,\Phi}/20$, but even
in this case the remaining non-asymmetric density will be negligibly
small~\cite{Acharya:2009zt}.

The role of the hidden $Z'$ in our model is to ensure the 
thermalization and symmetric annihilation of $Y$ and $\Phi$.
A more minimal alternative is to couple $\Phi$ to the SM Higgs boson $h$ via the
operator $\xi\,|h|^2|\Phi|^2$.  For $\xi \gtrsim 10^{-3}$, this interaction,
together with $YX\Phi$ and $|\Phi|^4$, appears to be sufficient
for both thermalization and symmetric annihilation.

  The residual CP-asymmetric density of $Y,\Phi$ is not eliminated and
makes up the DM~\cite{Kaplan:2009ag}.  The relic number density is
fixed by total baryon number conservation: $n_Y = n_\Phi = n_B$.
Thus the ratio between the energy
densities of DM and visible baryons is
\be
\Omega_{\textrm{d}}/\Omega_b = (m_Y + m_\Phi)/m_p\; . \label{eq:omega}
\ee
Present cosmological observations imply $\Omega_d/\Omega_b = 4.97 \pm  
0.28$ \cite{Komatsu:2008hk}, which
corresponds to a range \mbox{$4.4 \gev \lesssim m_Y + m_\Phi \lesssim  
4.9 \gev$}, or
$1.7 \gev \lesssim m_Y, m_\Phi \lesssim 2.9 \gev$ when combined with the constraint $|m_Y - m_\Phi| < m_p + m_e$.

\noindent {\bf{III. Dark Matter Signatures:}}
\begin{figure}[ttt]
\begin{center}
\includegraphics{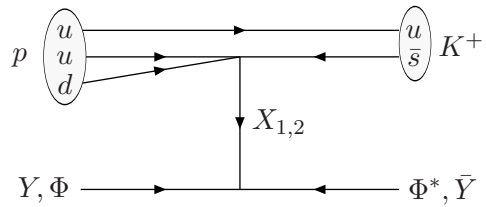} 
\end{center}
\caption{Diagram for induced nucleon decay processes $p Y \to K^+ \Phi^*$ and $p \Phi \to K^+ \bar{Y}$.}
\label{fig:indfeyn}
\vspace{-0.5cm}
\end{figure}
A novel signature of this mechanism is that DM
can annihilate nucleons through inelastic scattering processes 
of the form $Y N \to \Phi^* M$ and $\Phi N \to \bar{Y} M$ mediated by 
$X_{1,2}$, where $N$ is a nucleon and $M$ is a meson (Fig.~\ref{fig:indfeyn}).  
We call this process induced nucleon decay~(IND).  IND mimics
standard nucleon decay $N \to M \nu$, 
but with different kinematics of the daughter meson, summarized in 
Table~\ref{tab:ind}.  For down-scattering processes, where the mass 
of the initial DM state is greater than the final DM state, 
the meson momentum $p_M$ from IND can be much greater than 
in nucleon decay.  The quoted range of $p_M$ corresponds to 
the range of allowed masses $(m_Y,m_\Phi)$ consistent with 
Eqs.~(\ref{eq:massdiff}, \ref{eq:omega}).  For fixed masses, 
$p_M$ is monochromatic, with negligible broadening from the local 
DM halo velocity.  (We also note a related study considering lepton-number violating inelastic DM-nucleon scattering~\cite{Kile:2009nn}.)

\begin{table}[b!]
\begin{center}
\begin{tabular}{|c|c|c|}
\hline 
Decay mode & $p_M^{\textrm{SND}}$ (MeV) & $p_M^{\textrm{IND}}$ (MeV)  \\
\hline
$N \to \pi$ & 460 & 800 - 1400  \\
$N \to K$ & 340 & 680 - 1360 \\
$N \to \eta$ & 310 & 650 - 1340  \\
\hline
\end{tabular}
\end{center}
\caption{Daughter meson $M\in \{\pi, K, \eta\}$ momentum $p_M$ for standard nucleon decay (SND) and down-scattering IND.}
\label{tab:ind}
\vspace{-0.5cm}
\end{table}

To estimate the rate of IND we
consider the specific operator 
$(\lambda_a/M^2) (\bar{u}^cd)_R(\bar{X}s)_R$ that  
mediates $\Phi p \to \bar{Y} K^+$, illustrated in 
Fig.~\ref{fig:indfeyn}.  Treating the $\Phi$ and $Y$ 
states as spectators, the hadronic matrix element can be estimated 
from 
the value computed for the $p\to K^+\nu$ decay through 
the corresponding three-quark operator~\cite{Aoki:1999tw}.
We find that the sum of the IND scattering rates
$p\Phi \to K^+ \bar{Y} $ and $p Y\to K^+ \Phi^*$ is given by
\bea
(\sigma v)_{IND} = 
\mathcal{C}\, (10^{-39}cm^3/s)\;
\left|\sum_a\frac{\tev^3}{m_aM^2/\lambda_a^*\zeta_a}\right|^2
\label{eq:sigind}
\eea
where $0.5 < \mathcal{C} < 1.6$, depending on $m_{\Phi,Y}$ within 
the allowed range.
We expect IND modes from other operators to be roughly comparable.
This estimate, which relies on a chiral perturbation theory expansion
that is expected to be poorly convergent for $p_M \sim 1$ GeV, 
is approximate at best.

  An effective proton lifetime $\tau_p$ can be defined as the inverse IND 
scattering rate per target nucleon, $\tau^{-1}_p = n_{DM}(\sigma v)_{IND}$.
With a local DM density of $0.3\,\gev/cm^3$, 
$(\sigma v)_{IND} = 10^{-39}cm^3/s$ corresponds to a lifetime
of $\tau_p \simeq 10^{32}\,yr$.  This is similar to the current lifetime
bound on $p\to K^+ \nu$ of 
$2.3\times 10^{33}\,yr$~\cite{Kobayashi:2005pe}.  
However, existing nucleon decay bounds may not directly apply to IND due to 
the non-standard meson kinematics~\cite{paper2}, and additional suppression
can arise from the second factor in \eq{eq:sigind}.

  There is also a direct detection signal in our model due 
to the hidden $Z'$:  
$Y$ and $\Phi$ can scatter elastically off protons.  
The effective scattering cross-section per nucleon for either 
$Y$ or $\Phi$ is spin-independent and given by
\bea
\sigma_0^{SI} &=& (5\times 10^{-39}cm^2)\lrf{2Z}{A}^2\lrf{\mu_{N}}{\gev}^2
\\
&&~\times \lrf{e'}{0.05}^2
\lrf{\kappa}{10^{-5}}^2
\lrf{0.1\gev}{m_{Z'}}^4,
\nnmb
\eea
where $\mu_N$ is the DM-nucleon reduced mass. 
For a DM mass of $2.9\,\gev$, this is slightly below the current best 
limit from CRESST~\cite{Angloher:2002in}. 
The effective nucleon cross-section will be much lower if the
hidden vector is replaced by a $\Phi$-Higgs coupling.

  Annihilation of DM can generate energetic particles
that destroy the products of nucleosynthesis~\cite{Hisano:2009rc}, 
create a neutrino flux~\cite{Beacom:2006tt},
and modify the properties of the cosmic microwave 
background~\cite{Slatyer:2009yq}.  In our
scenario there is almost no direct DM annihilation after 
freeze-out, but similar effects can arise from IND scattering.  
However, the rate for this scattering given in \eq{eq:sigind} 
is many orders of magnitude less than the corresponding limits on
DM annihilation cross-sections~\cite{Hisano:2009rc,Slatyer:2009yq}, 
and thus we expect that IND will have no visible effect in the cosmological setting.

  The effects of IND can become important in astrophysical 
systems with very large densities of nucleons such as neutron stars
and white dwarfs.  In both cases, we find that the rate of IND in
the stellar core typically becomes large enough that it reaches a steady 
state with the rate of DM 
capture through elastic scattering~\cite{Kouvaris:2007ay,Bertone:2007ae}.  This has the effect of heating
the stellar interior in much the same way as DM annihilation.
Nucleons are also destroyed, but the number is only a tiny fraction 
of the total.
Current observations of white dwarfs constrain stellar heating 
by DM or IND, but the bounds depend on the local DM 
densities which are not known precisely~\cite{McCullough:2010ai,Hooper:2010es}.
These bounds are much weaker than for monopole catalysis of 
nucleon decay~\cite{Freese:1983hz} since the anti-DM product of IND is 
unable to destroy any more nucleons.



\noindent {\bf{IV. Conclusions:}}
  We have presented a novel mechanism to generate dark matter and baryon densities
simultaneously.  Decays of a massive $X_1$ state split baryon number
between SM quarks and antibaryons in a hidden sector.  These hidden
antibaryons constitute the dark matter.  An important signature of this
mechanism is the destruction of baryons by the scattering of hidden dark matter.

\noindent{\bf{Acknowledgements:}}  We thank M. Buckley, K. Freese, G. Kribs, M. Ramsey-Musolf, 
J. Shelton, A. Spray, M. Wise, and K. Zurek for helpful conversations.
DM and KS thank the Aspen Center for Physics and Perimeter 
Institute for Theoretical Physics for hospitality while this work was being 
completed.  ST thanks Caltech where a portion of this work was completed. 
The work of HD is supported in part by the United States Department of
Energy under Grant Contract DE-AC02-98CH10886.  The research of DM and KS 
is supported in part by NSERC of Canada Discovery Grants.

\end{document}